\documentclass[aps,showpacs,amsmath,amssymb]{revtex4}

\begin{document}
\title{Oscillations of a rapidly rotating annular Bose-Einstein
condensate}
\author{Alexander L. Fetter}
\email{fetter@stanford.edu}
\affiliation{Geballe Laboratory for Advanced Materials and Department of Physics, \\ Stanford University, Stanford, CA 94305-4045}
\date{\today}
 
 \begin{abstract}  
 
 A time-dependent variational Lagrangian analysis based on the
Gross-Pitaevskii energy functional serves to study the
    dynamics of a metastable giant vortex in a  rapidly
rotating Bose-Einstein condensate.  The resulting
oscillation frequencies of the core radius reproduce the trends seen in
recent experiments [Engels {\it et al.\/}, Phys.~Rev.~Lett.~{\bf 90},
170405 (2003)], but the theoretical values are smaller by a factor
$\approx 0.6$-0.8.
 \end{abstract}
 
 
 \maketitle
 
 \section{Introduction}
 
 One of the most characteristic features of neutral superfluids is their
inability to rotate uniformly with solid-body velocity ${\bf v}_{\rm
sb} = \bf {\Omega\times r}$.  Instead, a rotating  superfluid contains one
or more singly quantized vortex lines with circulation $h/M$, where $M$ is
the atomic mass. This phenomenon is familiar in  rotating superfluid
$^4$He~\cite{Donn91}, although visualization of the resulting vortex
arrays is quite difficult~\cite{Yarm79}.  The  creation of the first
Bose-Einstein condensate (BEC) in  dilute
$^{87}$Rb gas~\cite{Ande95}  challenged experimentalists to demonstrate
the existence of similar quantized vortices in   dilute BECs and to
study their properties.  After the development  of  techniques for
creating a single vortex~\cite{Matt99} and small vortex
arrays~\cite{Madi00},  rapid progress led to  large vortex
lattices~\cite{Abo01} and  rotating condensates with large
angular momentum~\cite{Halj01}.

One recent method to increase the angular momentum has used  an
intense focused laser beam to remove the central region of a rotating
condensate~\cite{Enge03,Enge03a}.  This method notes that the particles
near the center have relatively little angular  momentum, so that the
resulting annular condensate has an increased angular momentum per
particle.  In  this way, it has been feasible to reach rotation rates
$\Omega\approx 0.97\,
\omega_\perp$, where $\omega_\perp$ is the  radial confining trap
frequency.  This process creates a relatively long-lived  ``giant vortex
core'' that contains a large number of phase singularities, resulting in
a macroscopic circulation  around its boundary.  The remaining annular
fluid has a dense vortex lattice that rotates at the angular
velocity $\Omega$.  In addition, the radius of the  core oscillates at a
frequency of order 3-3.5 $\omega_\perp$, with relatively weak
damping~\cite{Enge03,Enge03a}.  Recently, a theoretical study of this
behavior has relied on numerical simulation of the time-dependent
Gross-Pitaevskii (GP) equation in two dimensions~\cite{Simu03} and found
an oscillation frequency $\sim 2.6
\,\omega_\perp$, somewhat smaller than the observed values.
The present work provides an alternative theoretical study of the
same  phenomenon using a variational Lagrangian that  contains the inner
radius and the amplitude of the radial velocity as time-dependent
parameters.  The resulting Lagrange's equations yield an undamped
oscillation with definite amplitude and a frequency that varies between 0.6
and 0.8 of various measured values~\cite{Enge03,Enge03a}.  Although the
two theoretical approaches appear quite different, they both rely on the
GP equation and  both predict oscillation frequencies that are definitely
smaller than the observed ones.

Section~{\ref{proc}} reviews the variational Lagrangian
procedure and applies it to the present problem. Sections~{\ref{2d}}
and {\ref{3d}} contain analyses of two specific models.  The first is a
quasi-two-dimensional Thomas-Fermi (TF) condensate with tight confinement
in the $z$ direction;  the two-dimensional character of the condensate
wave function simplifies the analysis considerably.  The second is a more
realistic three-dimensional TF condensate, but the final results are
rather similar in both cases.  The comparison with the experimental
observations is discussed in  Sec.~{\ref{obs}}.

\section{Basic procedure}\label{proc}

The  approach  here is the variational Lagrangian method, in which a
trial condensate wave function $\Psi$ includes various parameters whose
time dependence follows from  Lagrange's equations for the Lagrangian
\begin{equation}\label{lag}
L=\case{1}{2}i \hbar\int dV \,\left(\Psi^*\frac{\partial
\Psi}{\partial t}-\Psi\frac{\partial\Psi^*}{\partial t}\right) - E[\Psi],
\end{equation}
where $E[\Psi]$ is the energy functional. This flexible procedure has
served  successfully to study the low-lying collective modes of a
condensate~\cite{Pere96} and the precession of a single vortex
line~\cite{Svid00,Lund00,Fett01,Fett01a} in a trapped condensate. It is
convenient here to work in the laboratory frame of reference, and to
impose conservation of particle number and angular momentum explicitly.

The experiment~\cite{Enge03} studies $N\sim 10^6$ atoms of $^{87}$Rb in
an axisymmetric  trap with
$\omega_\perp/2\pi = 8.3
$ Hz and
$\omega_z/2\pi = 5.4 $ Hz, so that the relevant oscillator lengths are
     $d_\perp\approx 3.73\ \mu$m and $d_z\approx
4.62\ \mu$m.  Thus the nonrotating condensate is somewhat elongated, with
TF dimensions
$R_z(0)/R_\perp(0)= \omega_\perp/\omega_z \approx 1.54$.  When the
condensate rotates rapidly, however, it expands radially and contracts
axially according to~\cite{Fett01b}
\begin{equation}
\frac{R_\perp(\overline\Omega)}{R_\perp(0)} =
(1-\overline\Omega^2)^{-3/10},\quad \hbox{and}\quad
\frac{R_z(\overline\Omega)}{R_z(0)} =
(1-\overline\Omega^2)^{1/5},
\end{equation}
    where $\overline\Omega
=\Omega/\omega_\perp$.  Taking $\overline\Omega = 0.97$~\cite{Enge03},
     the standard TF expressions~\cite{Fett01} yield
$R_\perp(\overline
\Omega)/d_\perp
\approx 19.9$ and
$R_z(\overline\Omega)/d_\perp\approx 7.45$, with
$R_z(\overline\Omega)/R_\perp(\overline\Omega)=
(1-\overline\Omega^2)^{1/2}R_z(0)/R_\perp(0)\approx
0.37$.  Thus the rotating condensate is definitely disk shaped.

     In the TF limit, the gradient of the density $|\Psi|^2$ is small, and
the energy functional becomes
\begin{equation}
E[\Psi] \approx \int dV\,\left(\frac{1}{2}M v^2\,|\Psi|^2 + V_{\rm
tr}\,|\Psi|^2 +\frac{2\pi \hbar^2 a}{M}|\Psi|^4\right),
\end{equation}
where $v^2$ is the squared velocity of the superfluid, $V_{\rm tr}$ is the
trap potential, and
$a\approx 5.77$ nm is the
$s$-wave scattering length.  The density $|\Psi|^2$ is taken as the
familiar quadratic TF distribution with an empty core of radius
$r_c$
\begin{equation}\label{density}
|\Psi|^2 = |\Psi_{TF}|^2\theta(r-r_c).
\end{equation}
The core radius $r_c$ serves as one dynamical variable, and the resulting
radial  motion  requires a time-dependent radial velocity $v_r=
(\hbar/M)\,\partial S/\partial r$ in addition to the azimuthal component
$v_\phi = (\hbar/Mr)\,\partial S/\partial \phi$, where $S$ is the phase of
the condensate wave function.

\section{quasi-two-dimensional condensate}\label{2d}

To simplify the analysis, the
condensate is first assumed to be thin in the axial direction  with a
Gaussian dependence, so that $\Psi({\bf r},z) \approx \Psi_{\rm 2d}({\bf
r})\,
\psi_0(z)$.  Here,   $\psi_0(z) = (\sqrt\pi d_z)^{-1/2}
\exp(-\case{1}{2}z^2/d_z^2)$, where $d_z = \sqrt{\hbar/M\omega_z}$.   For
the two-dimensional wave function $\Psi_{\rm 2d}({\bf r})$, it is
convenient to use dimensionless variables with frequencies (and inverse
time) scaled by the radial trap frequency
$\omega_\perp$, energies scaled by $\hbar\omega_\perp$,  distances
scaled by the radial oscillator length
$d_\perp= \sqrt{\hbar/M\omega_\perp}$, and the condensate wave function
$\psi=\Psi/\sqrt N$ normalized to 1.

The trial wave function used here  corresponds to a radial TF
two-dimensional density for $r\le R$ with the central region  of radius
$r_c$ removed, so that
\begin{equation}
|\psi_{\rm 2d}(r)|^2 = C_{\rm
2d}^2\left(1-\frac{r^2}{R^2}\right)\,\theta(R-r)\,\theta(r-r_c),
\end{equation}
where $R=R_\perp/d_\perp$ is now dimensionless and $C_{\rm 2d}$ is a
normalization  constant.  The phase $S$ of the wave function is assumed
to have the radial part
\begin{equation}\label{radial}
S_r= \alpha\,(\case{1}{2} r^2-rR)
\end{equation}
     to allow for the oscillatory radial motion,  where $\alpha$ is a
    time-dependent parameter.  In addition, the phase has the familiar
contribution $S_v$ from the vortices plus the
circulation $\nu_0$ around the hole representing the missing vortices
\begin{equation}
S_v({\bf r}) =\sum_j\arctan\left(\frac{y-y_j}{x-x_j}\right) +
\nu_0 \phi,
\end{equation}
where the sum is over all vortices  $j$.

The radial velocity $v_r=\partial S_r/\partial r = \alpha \,(r-R) $
vanishes at the outer boundary, which remains fixed in this model. For
simplicity,  only
$r_c$ and $\alpha$ are taken as dynamical variables.
    In contrast,  the vortices  are {\it
not\/} treated as dynamical variables; instead, they are assumed to have
uniform dimensionless areal density
$\rho$ and to move with the local azimuthal velocity.  In  detail, the
azimuthal component of the dimensionless velocity  is
\begin{equation}
v_\phi = \sum_j \frac{r-\hat r\cdot {\bf r}_j}{|{\bf r-r}_j|^2}
+\frac{\nu_0 }{r},
\end{equation} where $\nu_0 =
\pi
\rho_0 r_0^2$ is the constant quantum number of the circulation  around
the  hole of initial radius
$r_0$ with initial vortex density  $\rho_0$.  The sum can be
approximated by an integral
\begin{eqnarray}\label{azim}
v_\phi &\approx & \rho \int d^2 r'
\frac{r-r'\cos\phi'}{r^2-2rr'\cos\phi' + r'^2}
\nonumber  + \frac{\nu_0}{r} = \frac{2\pi \rho}{r}\int_{r_c}^r r'dr' +
\frac{\nu_0}{r} \\
    & = & \pi \rho
\,r-\pi\rho\,\frac{r_c^2}{r}  +\frac{\nu_0}{r}.
\end{eqnarray}
      Note that
    the initial azimuthal component of
velocity
$\pi
\rho_0r $ represents a pure solid-body  rotation at  a dimensionless
angular velocity $\overline \Omega = \pi \rho_0$ since the last two
terms of Eq.~(\ref{azim}) cancel for $r_c = r_0$ (when $\rho = \rho_0$).

The analysis is
carried out in the laboratory frame, and the conservation  of particle
number and angular momentum are explicitly incorporated into the resulting
Lagrangian. It is straightforward to obtain the normalization constant
\begin{equation}\label{norm}
C_{\rm 2d}^2 = \frac{2}{\pi R^2\,(1-y^2)^2},
\end{equation}
where $y \equiv r_c/R$  determines the instantaneous  position  of the
core radius.  Similarly, the angular  momentum per particle (in units of
$\hbar$) is
\begin{equation}\label{dens}
l_{\rm 2d} = \int d^2 r\,rv_\phi\,|\psi_{\rm 2d}|^2 = \case{1}{3} \pi
\rho\, R^2 (1-y^2) +
\nu_0 =
\case{1}{3}
\pi
\rho_0\,R^2(1+ 2y_0^2),
\end{equation}
expressed in terms of  the initial parameters $\rho_0$ and $y_0$.
Since $l_{\rm 2d}$ and $\nu_0$ are taken  as constants, this relation
     determines the actual vortex density
\begin{equation}
\rho =\frac{3}{\pi R^2}\,\frac{l_{\rm 2d}-\nu_0}{1-y^2}
\end{equation}
as the  core
radius $y$ changes with  time.  Note that the total number of vortices
$\pi \rho R^2(1-y^2)= \pi \rho (R^2-r_c^2)$ in the annular region $r_c\le
r\le R$ remains
    constant, so that the vortex density changes only
because the inner radius $r_c$ changes. Correspondingly, the initial
conditions fix the quantity
$l_{\rm 2d}-\nu_0 = \case{1}{3}R^2\overline \Omega (1-y_0^2)$.  Assuming
typical values
$R\approx 20$,
$\overline\Omega\approx 0.97$, and $y_0\approx 1/3$, I find $l_{\rm 2d}
\approx 158$ and $\nu_0\approx 43$.

     After  the $z$ integration is carried out with the tightly
confined Gaussian, the dimensionless Lagrangian per particle becomes
\begin{equation}\label{lagr}
L_{\rm 2d} = \case{1}{2} i\int d^2 r \,\bigg(\psi_{\rm 2d}^*\frac{\partial
\psi_{\rm 2d}}{\partial t} -\psi_{\rm 2d}\frac{\partial\psi_{\rm
2d}^*}{\partial t} \bigg) - E_{\rm 2d}[\psi_{\rm 2d}],
\end{equation}
where
\begin{equation}\label{en}
     E_{\rm 2d}[\psi_{\rm 2d}] = \int d^2r\,\bigg[\case{1}{2}  \big(v_r^2
+v_\phi^2 \big)\,|\psi_{\rm 2d}|^2  +
\case{1}{2}\, r^2|\psi_{\rm 2d}|^2 +
\sqrt{2\pi}\,\frac{Na}{d_z}\,|\psi_{\rm 2d}|^4\bigg]
\end{equation}
is the two-dimensional energy functional.   All the integrals
can be performed analytically and Eqs.~(\ref{norm}) and (\ref{dens})
     ensure the
conservation of particles and angular momentum.  Note that the same
Lagrangian also describes a condensate that is uniform along a length $L$
in the $z$ direction. In this case, the coefficient of the quartic term
    in $E_{\rm 2d}$ is changed to $2\pi Na/L$, where $N/L$ is the number of
particles  per unit length.

The resulting Lagrangian
depends on  two dynamical variables: $\alpha$ that fixes the radial
component of velocity and $y \equiv r_c/R$ that fixes the position of the
inner core radius.  It is convenient to use the equivalent quantity
\begin{equation}\label{calL}
{\cal L}_{\rm 2d} = -L_{\rm 2d}/R^2= \dot \alpha \big[f_{\rm 2d}(y) -
\case {1}{2}
\big] +
\alpha^2f_{\rm 2d}(y) +{\cal E}_{\rm 2d}(y)
\end{equation}
where the first term arises from the explicit time derivative  in
(\ref{lagr}), the second is the radial kinetic energy, and ${\cal
E}_{2d}$ constitutes the remaining terms in the energy.  It has the
explicit form
\begin{eqnarray}
{\cal E}_{\rm 2d}(y) &= &
     \overline\Omega^2\,(1-y_0^2)^2\,\bigg[g_{\rm 2d}(y) +
\frac{h(y)}{1-y^2}\, +  h(y)^2j_{\rm 2d}(y)\bigg]\nonumber\\
& & +\case{1}{6}(1+2y^2) + \frac{4\sqrt 2}{3\sqrt
\pi}\,\frac{Na}{d_z R^4}\,\frac{1}{1-y^2},\label{calE}
\end{eqnarray}
where the first term (proportional to $\overline\Omega^2$) is the
azimuthal kinetic energy, the second is the trap energy, and the last is
the interaction energy (note that this term is typically small, despite
the large value of $Na/d_z\sim $ a few thousand, owing to the factor
$R^{-4}\sim 6\times 10^{-6}$). To be very specific, ${\cal L}_{\rm 2d}$
involves the following functions
\begin{eqnarray}
f_{\rm 2d}(y) &=& \frac{(1-y)^2(2+8y+5y^2)}{15(1+y)^2},\\
g_{\rm 2d}(y)&=&\frac{1+2y^2}{6(1-y^2)^2},\\
h(y)&=& \frac{1}{1-y_0^2}-\frac{1}{1-y^2},\\
j_{\rm 2d}(y)&=&
\frac{2\ln(1/y)}{(1-y^2)^2}-\frac{1}{1-y^2}.
\end{eqnarray}

\section{three-dimensional  Thomas-fermi condensate}\label{3d}

Before studying Lagrange's equations for  ${\cal L}_{\rm 2d}$, it is
convenient to consider the corresponding case of a
three-dimensional Thomas-Fermi condensate,
where
\begin{equation}
|\psi_{TF}(r,z)|^2 = C_{TF}^2
\left(1-\frac{r^2}{R^2}-\frac{z^2}{R_z^2}\right)\,
\theta\left(1-\frac{r^2}{R^2}-\frac{z^2}{R_z^2}\right)\,\theta(r-r_c),
\end{equation}
using the same scale factors $\omega_\perp$ and $d_\perp$ to define
dimensionless variables. The normalization  integral gives
\begin{equation}\label{normTF}
C_{TF}^2 = \frac{15}{8\pi R^2R_z\,(1-y^2)^{5/2}},
\end{equation}
which can be compared to Eq.~(\ref{norm}).  Similarly, the angular
momentum becomes
\begin{equation}
l_{TF} = \case{2}{7}\pi\rho R^2( 1 - y^2) + \nu_0,
\end{equation}
and the conservation of angular momentum then yields the vortex density
\begin{equation}
\rho = \frac{7}{2\pi R^2}\,\frac{l_{TF}-\nu_0}{1-y^2}.
\end{equation}

All the terms in the Lagrangian  can be evaluated
analytically, and the resulting expression has the same form as
Eq.~(\ref{calL}) but with  different functions
\begin{equation}\label{calLtf}
{\cal L}_{TF} = \dot \alpha \big[f_{TF}(y) -
\case {1}{2}
\big] +
\alpha^2f_{TF}(y) +{\cal E}_{TF}(y).
\end{equation}
     In particular,
\begin{equation}
f_{TF}(y) =  \frac{9+5y^2}{14} -\frac{5}{32}\,\frac{\pi
+\case{2}{3}y\,(1-y^2)^{1/2} (3 - 14 y^2+8 y^4) -2\arcsin
y}{(1-y^2)^{5/2}}
\end{equation}
     and
\begin{eqnarray}
{\cal E}_{TF}(y) &= &
     \overline\Omega^2\,(1-y_0^2)^2\,\bigg[g_{TF}(y) +
\frac{h(y)}{1-y^2}\, +  h(y)^2j_{TF}(y)\bigg]+ \nonumber\\
& &\frac{1}{14}\,[2+5y^2
     +(1-\overline \Omega^2)(1-y^2)] +\frac{15}{7} \frac
{Na}{d_\perp}\,\frac{1}{R^4R_z}\,\frac{1}{(1-y^2)^{3/2}},\label{calEtf}
\end{eqnarray}
The extra factor of $R_z^{-1}$ in the  interaction term
     will be seen to
eliminate  the dependence of the number of atoms (assuming only that the
condensate remains in the TF limit). In  addition, the remaining functions
have the altered forms
\begin{eqnarray}
     g_{TF}(y)& =&
\frac{2+5y^2}{14\,(1-y^2)^2},\\
j_{TF}(y) & = & \frac{5}{6} \,\frac{(1-y^2)^{1/2}\,(-4+y^2) -3\ln y +
3\ln[1+(1-y^2)^{1/2}]}{(1-y^2)^{5/2}}.
\end{eqnarray}

\section{Discussion of resulting motion}\label{obs}

Variation of the effective Lagrangian ${\cal L}$
     with respect to
$\alpha$ yields the dynamical equation
\begin{equation}\label{alphaeq}
\dot y \,\frac{df(y)}{dy}= 2f(y) \,\alpha,
\end{equation}
which expresses $\alpha$ in terms of $y$ and $\dot y$.  Similarly,
variation with respect to $y$ yields a more complicated equation
\begin{equation}\label{alphadot}
\frac{\partial {\cal L}}{\partial y} =
\left(\dot\alpha+\alpha^2\right)\frac{df(y)}{dy} +
\frac{d{\cal E}(y)}{dy} = 0
\end{equation}
that
expresses
$\dot
\alpha$ and $\alpha$ in terms of $y$.

Unfortunately, it is not easy to
interpret this pair of coupled dynamical equations, and it is preferable
to assume small oscillations about the initial conditions $y(0) = y_0$
and $\dot y(0) = 0$ (namely the hole in the rapidly rotating
condensate  starts from rest). Consequently, $\alpha$ itself is a small
quantity, as is $\delta = y_0 - y$, which is positive when
the hole contracts from its initial scaled radius $y_0$.  In this way,
Eq.~(\ref{alphaeq}) becomes
\begin{equation}
\dot\delta = - \frac{2f(y_0)}{f'(y_0)}\,\alpha,
\end{equation}
where the prime denotes a derivative with respect to $y$.  Similarly,
Eq.~(\ref{alphadot}) has the linearized form
\begin{equation}
\dot \alpha = \frac{-{\cal E}'(y_0) +\delta\, {\cal E}''(y_0)}{f'(y_0)}
\end{equation}
A combination  of these two linearized equations yields  a
simple dynamical equation
\begin{equation}\label{osc}
\ddot\delta +\omega^2\,\delta = \omega^2 A,
\end{equation}
     where
\begin{equation}
\omega^2 = \frac{2 {\cal E}''(y_0)\,f(y_0)}{\left[f'(y_0)\right]^2},
\end{equation} and
\begin{equation}
A = \frac{{\cal E}'(y_0)}{{\cal E}''(y_0)}.
\end{equation}
     They depend on the initial radius $y_0 = r_0/R$ and
on the small parameters $Na/d_zR^4$ in the two-dimensional case and
$Na/d_\perp R_zR^4$ for the three-dimensional TF case (these latter
contributions come from the interaction energy, which is small because
the expanded outer radius reduces the particle density far below its
value for a nonrotating condensate).

Equation (\ref{osc})  has the
solution
\begin{equation}
\delta(t) =  A(1-\cos\omega t).
\end{equation}
It represents an inward oscillation of the core radius with angular
frequency $\omega$ (measured in units of $\omega_\perp$) and amplitude $ A
$ (like $y$,
$A$ is measured in units of $R=R_\perp/d_\perp$). As a result,
conservation of angular momentum and the
parabolic trap potential dominate  the dynamical motion because the
interaction  term  makes only a small contribution; this is the effect of
the Coriolis force mentioned in
\cite{Enge03}.  In all  cases, the amplitude
$A$ is relatively  small, partially justifying the expansion through
quadratic contributions to ${\cal L}$.  Note, however, that the
minimum core size is $r_0-2A R$, which corresponds to a significant
shrinkage for the smaller rotation rates $\overline \Omega\lesssim 0.9$.

Table I  contains typical calculated values for some of the experimental
parameters used in \cite{Enge03}, using the functions for a
quasi-two-dimensional condensate.  Since the present theory is
rather crude, however, the general three-dimensional TF radii have been
used even  for the  rotating quasi-two-dimensional condensate.
The  choice  $N=2.2\times 10^6$ corresponds to  Fig.~4 of
Ref.~\cite{Enge03}, which  shows  a rotating condensate with
$\overline
\Omega = 0.9$;  the core executes weakly damped oscillations with
dimensionless frequency
$\omega/\omega_\perp = 3.5$. This frequency is higher than the value
$\omega/\omega_\perp = 2.32 $ found here for a core radius $1/3$ of the
condensate radius, but the dependence of the oscillation  frequency and
amplitude on the rotation frequency
$\overline\Omega$ both show the observed trend (namely, ``decreasing the
initial rotation rate of the condensate leads to faster core-oscillation
frequencies and increases the amplitude of this
oscillation"~\cite{Enge03}).  The corresponding values for $N=3.0\times
10^6$ illustrate the extremely weak dependence on the number of atoms.
In fact, if strict two-dimensional TF radii were used, the dependence on
$N$ would cancel completely, as will be seen to hold for the
three-dimensional TF condensate.

Table II is similar to Table I except that it is
evaluated for the three-dimensional Thomas-Fermi condensate density.  In
this case, it is not difficult to use the standard TF expressions for the
condensate radii~\cite{Fett01} to verify that the ratio $(15Na/d_\perp)
(R^4R_z)^{-1}$ in fact reduces to the value
$1-\overline\Omega$, so that the $N$ dependence indeed drops out.  In all
the cases considered, the three-dimensional TF condensate yields
somewhat lower oscillation frequencies and smaller amplitude for the
motion.

To provide a more detailed comparison between the theoretical predictions
and experimental values, Table III contains 5 sets of measured
frequencies~\cite{Enge03a}  along with the corresponding predictions of
both  models.   For the quasi-two-dimensional condensate, the
ratio of predicted value $\omega_{\rm th}$ to measured value
$\omega_{\rm obs} $  varies between 0.61 and 0.79, whereas for the
three-dimensional TF condensate, the same ratio varies between 0.49 and
0.76.  In both cases, the agreement improves considerably at the fastest
rotations.  This trend may reflect the present use of a spatially averaged
velocity for that induced by the vortices and the macroscopic
circulation, ignoring the discrete character of the vortices; such an
approximation should work best at large $\overline \Omega$.

One possible explanation for the
     discrepancy between the experimental values and the results of  the
present variational Lagrangian formalism  is that the actual dynamical
condensate does not conserve the number of vortices; in this case,
angular momentum would not strictly be conserved.  At present, it is not
clear  how to incorporate such additional dynamical variables into the
analysis.  In any case, it is notable that the numerical simulation  of
Simula {\it et al.\/}~\cite{Simu03} also seems to predict an oscillation
frequency that is too small by a comparable amount.  Conceivably, the
quite dramatic perturbation caused by the intense laser beam in burning a
relatively large hole may well produce significant depletion, so that a
theory based solely on the time-dependent GP equation would need to be
modified to include noncondensate contributions.

\acknowledgments

I thank P. Engels for providing me with the data sets used in
Table III, and  both  P. Engels and V. Schweikhard for valuable
discussions and comments.  I am grateful for the hospitality of the Aspen Center for
Physics, where part of this work originated.

 \begin{table}
\caption{Calculated parameters for a quasi-two-dimensional condensate}
\begin{ruledtabular}
\begin{tabular}{ccc|cc|cc}\noalign{\vspace{.1cm}}

\multicolumn{3}{c}{} 
&\multicolumn{2}{c}{$N=2.2\times 10^6$}
&\multicolumn{2}{c}{$N=3.0\times 10^6$} \\ \noalign{\vspace{.1cm}}
  \hline   \noalign{\vspace{.1cm}}
$r_0/ R$&  $\overline \Omega = \Omega / \omega_\perp $ 
& $R=R_\perp / d_\perp$ &
$\omega / \omega_\perp$ &
$A$ & $\omega/\omega_\perp$ &
$A$ \\ \noalign{\vspace{.1cm}}
 \hline \noalign{\vspace{.1cm}}
     1/3 & 0.90 & 13.20 & 2.32 & 0.097  & 2.34 & 0.099\\
     1/3 & 0.94 & 15.29 & 2.19 & 0.063  & 2.20 & 0.064\\
     1/3 & 0.97 & 18.74 & 2.10 & 0.032  & 2.10 & 0.033\\
\noalign{\vspace{.1cm}}
\hline
\noalign{\vspace{.1cm}}
     1/5 & 0.90 & 13.20 & 2.50 & 0.081  & 2.52 & 0.083\\
     1/5 & 0.94 & 15.29 & 2.32 & 0.054  & 2.33 & 0.056\\
     1/5 & 0.97 & 18.74 & 2.18 & 0.029  & 2.19 & 0.030
\end{tabular}
\end{ruledtabular}
\end{table}

\begin{table}
\caption{Calculated parameters for a three-dimensional  TF condensate (the
predicted frequencies and amplitudes are independent of $N$ as long as
the TF approximation remains valid)}
\vspace{.2cm}
\begin{ruledtabular}
\begin{tabular}{ccc|cc|cc}
\noalign{\vspace{.1cm}}
\multicolumn{3}{c|}{}
&\multicolumn{2}{c|}{$r_0/R = 1/3$}&\multicolumn{2}{c}{$r_0/R= 1/5$}\\
     \noalign{\vspace{.1cm}}
\hline \noalign{\vspace{.1cm}}
     $\overline\Omega=\Omega/\omega_\perp$ & $
R=R_\perp/d_\perp$ &$R_z/d_\perp$ &
$\omega/\omega_\perp$  &
$A$ &
$\omega/\omega_\perp$  &
$A$ \\ \noalign{\vspace{.1cm}}
\hline \noalign{\vspace{.1cm}}
     0.90 & 13.20 & 8.85 & 2.09 & 0.061  & 2.20 & 0.052 \\
     0.94 & 15.29 & 8.02 & 2.06 & 0.038  &  2.15 & 0.034 \\
     0.97 & 18.74 & 7.00 & 2.04 & 0.020 &2.11 & 0.018
\end{tabular}
\end{ruledtabular}
\end{table}

\begin{table}
\caption{Comparison of selected experimental values with theoretical
predictions for two-dimensional and three-dimensional models}
\vspace{.2cm}
\begin{ruledtabular}
\begin{tabular}{cccc|ccc|ccc}
\noalign{\vspace{.1cm}}
\multicolumn{4}{c|}{experimental values\tablenotemark[1]}
&\multicolumn{3}{c|}{quasi-two-dimensional
model}&\multicolumn{3}{c}{three-dimensional TF model}\\
     \noalign{\vspace{.1cm}}
\hline \noalign{\vspace{.1cm}}
     $r_0/R$\tablenotemark[2] & $\overline\Omega=\Omega/\omega_\perp$ & $N $
     & $\omega_{\rm  obs}/\omega_\perp$\tablenotemark[3]&
$\omega_{\rm th}/\omega_\perp$ & $\omega_{\rm th}/\omega_{\rm  obs}$ &
$A$  &$\omega_{\rm th} /\omega_\perp$  & $\omega_{\rm th}/\omega_{\rm
obs}$ &
$A$ \\ \noalign{\vspace{.1cm}}
\hline \noalign{\vspace{.1cm}}
0.40 & 0.79 & $2.8\times 10^6$ & 4.5 & 2.74 & 0.61  &
0.171  & 2.17 & 0.49& 0.122\\
     0.24 & 0.88 & $2.3\times 10^6$ & 4.0 & 2.48 & 0.62 & 0.098 & 2.16 & 0.54
& 0.064\\
     0.32 & 0.90 &  $2.1\times 10^6$ & 3.5 & 2.32 & 0.67  & 0.095  & 2.09 &
0.60 & 0.060\\
     0.32 & 0.95  & $2.5\times 10^6$ & 2.9 & 2.17 & 0.75 & 0.053  & 2.06 &
0.71& 0.032\\
     0.35 & 0.96 & $2.3\times 10^6$ & 2.7  & 2.13 & 0.79 & 0.044  & 2.05 &
0.76 & 0.027

\end{tabular}
\end{ruledtabular}
\tablenotetext[1]{Reference~\cite{Enge03a}.}
\tablenotetext[2]{Reference~\cite{Enge03a} assigns an uncertainty of
$\sim 20\%$ to the initial core radius $r_0$.}
\tablenotetext[3]{Reference~\cite{Enge03a} assigns an uncertainty of 6 Hz
to the first value of $\omega_{\rm obs} $ (37 Hz), namely $\sim 16\%$.
In contrast, the other experimental oscillation frequencies have
uncertainties of
$\le 1$ Hz, which is $\le 4\%$ even for the last value.}
\end{table}
 
 \end{document}